\documentclass{article}
\usepackage{graphicx} % Required for inserting images
\usepackage[a4paper, portrait, margin=1in]{geometry}
\usepackage{amsmath}
\usepackage{amssymb}
\usepackage{hyperref}
\usepackage{ mathrsfs }

\title{\large{\bf{Comments on \textit{``Orbits of particles with magnetic dipole moment around magnetized Schwarzschild
black holes: Applications to S2 star orbit"},  arXiv:2406.03371v2}}}

\author{Miles Angelo P. Sodejana}
\author{{Miles Angelo P. Sodejana } \\ {Department of Physics, University of Southern Mindanao } \\ {9407 Kabacan, Cotabato, Philippines} \\ {Email: mapsodejana@usm.edu.ph}}
\date{December 2024}

\begin{document}
\maketitle

\begin{abstract}
We provided comments on the article \textit{“Orbits of particles with magnetic dipole moment around magnetized Schwarzschild black holes: Applications to S2 star orbit”} by Uktamjon Uktamov, Mohsen Fathi, Javlon Rayimbaev, and Ahmadjon Abdujabbarov, from reference \cite{Uktamov:2024zmj}.  We derived the Hamilton-Jacobi equation used in the article from the Lagrangian utilized and found inconsistencies in the use of the interaction term which describes the magnetic dipole moment of the particle.  Consequently, this results in incorrect equations of motion for the particle.

\end{abstract}

\section{Hamilton-Jacobi Equation for Polarized Particles in Curved Spacetime}

We start by the dynamic Lagrangian for a polarized charged particle was given in reference \cite{Preti:2004xu} as
\begin{align}
    \mathcal{L} = \frac{1}{2} \left[(m+\mathcal{U}) g_{\mu \nu}\dot x^\mu \dot x^\nu - k \mathcal{U}\right], \label{1}
\end{align}
where $\mathcal{U}$ is a velocity-independent interaction term.  This gives us the generalized (canonical) momenta
\begin{align}
    p_\mu = \frac{\partial \mathcal{L}}{\partial \dot x^\mu } =(m+\mathcal{U}) g_{\mu \nu} \dot x^\nu. \label{2}
\end{align}Taking the Legendre transform of equation \eqref{1} and using \eqref{2}, we find the Hamiltonian of the system to be
\begin{align}
    \mathcal{H} &= p_\mu \dot x^\mu - \mathcal{L} =  p_\mu \dot x^\mu - \frac{1}{2} \left[(m+\mathcal{U}) g_{\mu \nu}\dot x^\mu \dot x^\nu - k \mathcal{U}\right] \notag \\ &=  (m+\mathcal{U})g_{\mu \nu}\dot x^\mu \dot x^\nu  - \frac{1}{2} \left[(m+\mathcal{U}) g_{\mu \nu}\dot x^\mu \dot x^\nu - k \mathcal{U}\right] \notag \\ 
    &= \frac{1}{2} \left(m + \mathcal{U}\right) g_{\mu \nu} \dot x^\mu x^\nu + \frac{1}{2}k\mathcal{U} \notag \\
    &=\frac{1}{2} \left(\frac{1}{m+ \mathcal{U}}g^{\mu \nu} p_\mu p_\nu + k\mathcal{U}\right). \label{3}
\end{align}If $\dot x^\mu = \frac{dx^\mu}{d\lambda} = \frac{dx^\mu}{d\tau}$ where $\tau$ denotes proper time, we require $\mathcal{H} = -m/2$ for a timelike worldline, which also corresponds to the velocity normalization condition $g_{\mu \nu} \dot x^\mu \dot x^\nu = -1$, where it implies from \eqref{3} that $k = 1$.  This gives us the mass-shell condition
\begin{align}
    g^{\mu \nu} p_\mu p_\nu = -\left(m+\mathcal{U}\right)^2. \label{4}
\end{align}Albeit it can easily be seen from \eqref{4}, we will show that this corresponds to the correct Hamilton-Jacobi equation of the system. By performing a canonical transformation from the phase space coordinates $(x^\mu, p_\mu)$ with a Hamiltonian $\mathcal{H}$ to the coordinates $(X^\mu, P_\mu)$ with a Hamiltonian $\mathcal{K}$, we get 
\begin{align}
    p_\mu \dot x^\mu - \mathcal{H} = P_\mu \dot X^\mu - \mathcal{K} + \dot{\mathcal{F}} \label{5}
\end{align}where $\mathcal{F}$ is a function of the phase space coordinates and the proper time $\tau$ such that $\frac{d x^\mu}{d\tau} = \dot x^\mu$. Using the transformations
\begin{align}
    p_\mu = p_\mu\left(x^\alpha, P_\nu, \tau\right) , \hspace{8mm} X^\mu = X^\mu\left(x^\alpha, P_\nu, \tau\right) \label{6}
\end{align}we get
\begin{align}
    p_\mu dx^\mu - \mathcal{H} d\tau = -X^\mu dP_\mu - \mathcal{K} d\tau + d\left(\mathcal{F}+ X^{\mu}P_{\mu}\right). \label{7}
\end{align}Defining the generating function $\mathcal{S}$ as
\begin{align}
    \mathcal{S}\left(x^{\alpha}, P_\mu, \tau\right) = \mathcal{F} + X^\mu P_\mu \label{8}
\end{align}we obtain
\begin{align}
    d\mathcal{S} = \frac{\partial \mathcal{S}}{\partial x^\mu} dx^\mu + \frac{\partial \mathcal{S}}{\partial P^\mu} dP^\mu + \frac{\partial \mathcal{S}}{\partial \tau} d\tau. \label{9}
\end{align}Equation \eqref{7} then gives 
\begin{align}
    \left(p_\mu - \frac{\partial \mathcal{S}}{\partial x^\mu}\right)dx^\mu - \left(\mathcal{H} + \frac{\partial \mathcal{S}}{\partial \tau}\right) = - \left(X^\mu - \frac{\partial \mathcal{S}}{\partial P^\mu}\right) dP^\mu - \mathcal{K} d\tau, \label{10} 
\end{align}which results to 
\begin{align}
    p_\mu = \frac{\partial \mathcal{S}}{\partial x^\mu}, \hspace{5mm} X^\mu = \frac{\partial \mathcal{S}}{\partial P^\mu}, \hspace{5mm} \mathcal{H} + \frac{\partial \mathcal{S}}{\partial \tau} = \mathcal{K}. \label{11}
\end{align}Setting $\mathcal{K} = 0$ and using the mass-shell condition, we obtain the correct Hamilton-Jacobi equation \cite{ Preti:2004xu}
\begin{align}
     g^{\mu \nu} \frac{\partial \mathcal{S}}{\partial x^\mu}\frac{\partial \mathcal{S}}{\partial x^\nu}= - \left(m + \mathcal{U}\right)^2. \label{12}
\end{align} 
This differs from the Hamilton-Jacobi equation provided in reference \cite{Uktamov:2024zmj} as
\begin{align}
    g^{\mu \nu} \frac{\partial \mathcal{S}}{\partial x^\mu}\frac{\partial \mathcal{S}}{\partial x^\nu}= - m^2 \left(1 - \frac{\mathcal{U}}{m}\right)^2 \label{13}
\end{align}
which could be explained by how they defined the velocity-independent interaction term $\mathcal{U} = D^{\mu \nu}F_{\mu \nu}$, but nevertheless does not follow from the established Lagrangian \eqref{1}.  Note that \eqref{12} has already been derived in reference \cite{Preti:2004xu}.

\section{Equations of Motion of the Magnetized Particles}
The non-vanishing components of the orthonormal magnetic field are
\begin{align}
    B^{\hat i} = B_0 \left[\sqrt{f(r)} \sin \theta \delta^{\hat i}_{\hat \theta} + \cos\theta \delta^{\hat i}_{\hat r}\right]. \label{14}
\end{align}
According to reference \cite{Preti:2004xu}, for a polarized particle, $\mathcal{U}$ can be expressed as 
\begin{align}
    \mathcal{U} = -\frac{1}{2} D^{\mu \nu} F_{\mu \nu} \label{15}
\end{align} where $D^{\mu \nu} F_{\mu \nu}$ is a scalar product of the polarization tensor $D^{\mu \nu}$ and the electromagnetic field tensor $F_{\mu \nu}$, which describes the interaction between the magnetic dipole moment of the particle and the external magnetic field.  The polarization tensor $D^{\mu \nu}$ relates to the magnetic dipole moment by the expression \cite{deFelice:2003wx}:
\begin{align}
    D^{\alpha \beta} = -\frac{\varepsilon^{\alpha \beta \sigma \nu}}{\sqrt{-g}}w_{\sigma}\mu_\nu, \hspace{8mm} D^{\alpha \beta}w_\beta = 0, \label{16}
\end{align} where $\mu^\nu$ and $w^\nu$ are the four-dipole moment and the four-velocity vector of the particle measured by a proper observer. We assumed that the magnetic dipole moment direction is perpendicular to the equatorial plane and parallel to the magnetic field lines with the following components: $\mu^{\hat i} = (0, \, \mu^{\hat \theta}, \, 0) $. 
 The product of the polarization
and the electromagnetic tensors at ZAMO can then be found as \cite{deFelice:2003wx}
\begin{align}
    D^{\mu \nu} F_{\mu \nu} = 2\mu^{ \alpha} B_{\alpha } = 2\mu^{\hat \alpha} B_{\hat \alpha } = 2\mu B\sqrt{f(r)} \sin \theta. \label{17}
\end{align}Equations \eqref{15} and \eqref{17} then gives us $\mathcal{U} = -\mu Bf^{1/2} \sin \theta$. This contradicts the equation provided in reference \eqref{1} as $\mathcal{U} = \mu^{\hat \alpha} B_{\hat \alpha } = \mu B f^{1/2} \sin \theta$.  Working on the equatorial plane $(\theta = \pi/2)$ we find the following conjugate momenta from \eqref{2} as
\begin{align}
    p_t/m  &= -\left(1 - \beta f^{1/2}\right) f \dot t= - \mathcal{E}, \notag \\
    p_r/m &= \left(1 - \beta f^{1/2}\right) f^{-1}\dot r, \notag \\
    p_{\phi}/m &= \left(1 - \beta f^{1/2}\right) r^2 \dot \phi  = l \label{18}
\end{align}where $\beta = \mu B/m$, and $\mathcal{E}$ and $l$ are the specific energy and specific angular momentum of the particle in motion and are conserved quantities. This is consistent with reference \cite{Uktamov:2024zmj}.  Another inconsistency also arose from the use of the action $\mathcal{S} = - Et + L\phi + \mathcal{S}_r(r)$.  From the mass-shell condition \eqref{4} and the conjugate momenta \eqref{18}, we find
\begin{align}
    g^{tt} \mathcal{E}^2 + g_{rr} \left(1 - \beta f^{1/2}\right)^2 &\dot r^2 + g^{\phi \phi} l^2  = -\left(1 - \beta f^{1/2}\right)^2 \notag \\
    g_{rr}  \left(1 - \beta f^{1/2}\right)^2 \dot r^2  &= -g^{tt} \mathcal{E}^2 - \frac{l^2}{r^2} - \left(1 - \beta f^{1/2}\right)^2  \notag \\
    \left(1 - \beta f^{1/2}\right)^2 \dot r^2 & = \mathcal{E}^2 - f(r) \left[\frac{l^2}{r^2} + \left(1 - \beta f^{1/2}\right)^2\right] \notag \\
    \left(1 - \beta f^{1/2}\right)^2 \dot r^2 & = \mathcal{E}^2 - V_{\text{eff}} \label{19}
\end{align}where 
\begin{align}
    V_{\text{eff}} = f(r) \left[\frac{l^2}{r^2} + \left(1 - \beta f^{1/2}\right)^2\right] \label{20}
\end{align}which is consistent with reference \cite{Uktamov:2024zmj} except for the lack of the factor $\left(1 - \beta f^{1/2}\right)^2$ multiplied by $\dot r^2$.  This might have been derived from the action $\mathcal{S} = - Et + L\phi + \mathcal{S}_r(r)$ where $p_r = g_{rr} \dot r$.  These errors have also been committed by similar papers on magnetized particles around magnetized black holes, including those that are cited by reference \cite{Uktamov:2024zmj}.  We also see that using $\mathcal{E}$ and $l$ from equation \eqref{18} together with the velocity normalization condition for timelike worldline $g_{\mu \nu} \dot x^\mu \dot x^\nu = -1$, we get
\begin{align}
    g_{tt}\frac{\mathcal{E}^2}{g_{tt}^2\left(1 - \beta f^{1/2}\right)^2} + &g_{rr} \dot r^2 + g_{\phi \phi}\frac{l^2}{g_{\phi \phi}^2\left(1 - \beta f^{1/2}\right)^2} = -1 \notag \\ g^{tt} \mathcal{E}^2 & + g_{rr} \left(1 - \beta f^{1/2}\right)^2 \dot r^2 + \frac{l^2}{g_{\phi \phi}} = -\left(1 - \beta f^{1/2}\right)^2 \notag \\
    -f^{-1} \mathcal{E}^2 & + f^{-1} \left(1 - \beta f^{1/2}\right)^2 \dot r^2 + \frac{l^2}{r^2} = -\left(1 - \beta f^{1/2}\right)^2 \label{21}
\end{align}which gives us equation \eqref{19}.  Our two methods of deriving $\dot r$ are consistent, in contrast to reference \cite{Uktamov:2024zmj}.  

We can use these equations of motion to derive the angular trajectories or so-called "orbit equation" for the particles, as has been done in reference \cite{Uktamov:2024zmj}.  Using the equations for $\dot \phi$ and $\dot r$, we find
\begin{align}
    \left(\frac{dr}{d\phi}\right)^2 = \frac{\dot r^2}{\dot \phi^2} &= \frac{\frac{\mathcal{E}^2 - f(r) \left[\frac{l^2}{r^2} + \left(1 - \beta f^{1/2}\right)^2\right]}{\left(1 - \beta f^{1/2}\right)^2}}{\frac{l^2}{\left(1 - \beta f^{1/2}\right)^2 r^4}} \notag \\ &=
    \left\{\mathcal{E}^2 - f(r) \left[\frac{l^2}{r^2} + \left(1 - \beta f^{1/2}\right)^2\right]\right\}\times \frac{r^4}{l^2} \label{22}
\end{align}which is different from what was derived in reference \cite{Uktamov:2024zmj}, given as
\begin{align}
    \left(\frac{dr}{d\phi}\right)^2 = \left[\mathcal{E}^2 - f(r) \left(\frac{l^2}{r^2} + 1 - \beta f^{1/2}\right)\right]\times \frac{\left(1 + \beta f^{1/2}\right)^2 r^4}{l^2}.\label{23}
\end{align}We see that $ \left(\frac{dr}{d\phi}\right)^2$ is simpler in equation \eqref{22} than in equation \eqref{23}.  In the weak electromagnetic interaction limit $(\beta << 1)$, equation \eqref{22} in first order $\beta$ becomes
\begin{align}
    \left(\frac{dr}{d\phi}\right)^2 = \left[\mathcal{E}^2 - f(r) \left(\frac{l^2}{r^2} + 1 - 2\beta f^{1/2}\right)\right]\times \frac{r^4}{l^2}, \label{24}
\end{align}in contrast to reference \cite{Uktamov:2024zmj}
\begin{align}
    \left(\frac{dr}{d\phi}\right)^2 = \frac{r^4}{l^2}\left\{\mathcal{E}^2 \left(1+ 2\beta f^{1/2}\right) - f   \left[\beta f^{1/2} \left(1+\frac{2l^2}{r^2}\right)+1+\frac{l^2}{r^2}\right] \right\}. \label{25}
\end{align}
Setting $u = 1/r$, we find
\begin{align}
    \left(\frac{du}{d\phi}\right)^2 = \frac{1}{l^2} \left[\mathcal{E}^2 - \left(1-2Mu\right) \left(l^2 u^2 + 1 - 2\beta\sqrt{1-2Mu}\right)\right]. \label{26}
\end{align}Employing the approximation $\beta \sqrt{1-2Mu} \approx \beta - \beta Mu + \mathcal{O}(u^2)  $, we obtain
\begin{align}
    \left(\frac{du}{d\phi}\right)^2  = 2Mu^3 + \left(\frac{4\beta M^2 u^2}{l^2} - 1\right)u^2 + \frac{2M(1-3\beta)}{l^2}u + \frac{\left(\mathcal{E}^2 + 2\beta -1\right)}{l^2}. \label{27}
\end{align}Note that we get the right expression when $\beta = 0$ for the neutral Schwarzschild black hole (see, for example, reference \cite{hobson2006general}).  This is much simpler than what was found in reference \cite{Uktamov:2024zmj}.  Using equation \eqref{27}, maybe extend to higher order approximation for $\beta \sqrt{1-2Mu}$, we can follow the methods employed in reference \cite{Uktamov:2024zmj}, but with the correct equations of motion.

Lastly, we can find the correct radial velocity of the particles from our equations of motion, given as
\begin{align}
    \left(\frac{dr}{dt}\right)^2 = \frac{\dot r^2}{\dot t^2} & = \frac{\frac{\mathcal{E}^2 - f(r) \left(1 - \beta f^{1/2}\right)^2}{\left(1 - \beta f^{1/2}\right)^2}}{\frac{\mathcal{E}^2}{\left(1 - \beta f^{1/2}\right)^2 f^2 }} \notag \\ &= f^2 - \frac{f^3}{\mathcal{E}^2}\left(1 - \beta f^{1/2}\right)^2. \label{28}
\end{align}Similarly, we can follow the methods conducted in reference \cite{Uktamov:2024zmj} to analyze equation \eqref{28}.

\section{Conclusion}
This article focused on deriving the correct equations of motion from the dynamic Lagrangian of polarized particles established in reference \cite{Preti:2004xu}, as a response to reference \cite{Uktamov:2024zmj}, the subject of this comment.  We found that there are inconsistencies in the use of the Lagrangian and the Hamilton-Jacobi equation, and in consequence, errors in the equations of motion. This article did not deal with the analysis conducted in reference \cite{Uktamov:2024zmj}, but merely focused on the fundamental equations. This comment aims to provide constructive insights and clarify certain aspects of the discussed work. We highly encourage an open dialogue and welcome further discussion to address any misunderstandings that may arise.
\newpage
\bibliographystyle{ieeetr}
\bibliography{references}
\end{document}